\documentclass[3p,times,twocolumn]{elsarticle}
\usepackage{ecrc}
\volume{00}
\firstpage{1}
\journalname{Nuclear Physics B Proceedings Supplement}
\runauth{R. Aloisio}
\jid{nuphbp}
\jnltitlelogo{Nuclear Physics B Proceedings Supplement}
\usepackage{amssymb}
\usepackage[figuresright]{rotating}

\begin{document}
\begin{frontmatter}
\dochead{}
\title{Ultra High Energy Cosmic Rays: A Short Review}
\author{Roberto Aloisio}
\address{INAF - Osservatorio Astrofisico Arcetri, largo E. Fermi 5, I-50125 Firenze (Italy) \\
INFN - Laboratori Nazionali Gran Sasso, ss 17bis km 18+910, I-67100 L'Aquila (Italy)}

\begin{abstract}
We will review the main physical aspects of Ultra High Energy Cosmic Rays. We will discuss in particular 
their propagation through astrophysical backgrounds, focusing on the latest experimental 
observations of HiRes, Telescope Array and Auger. We will also review the issue of the transition between 
galactic and extra-galactic cosmic rays.
\end{abstract}

\begin{keyword}
Cosmic Rays Theory \sep  Cosmic Rays Observations \sep Particles Astrophysics \sep Astrophysical Backgrounds
\end{keyword}

\end{frontmatter}

\section{Introduction}
\label{intro}

Ultra High Energy Cosmic Rays (UHECR) are the most energetic particles observed in Nature, 
with energies up to few$\times 10^{20}$ eV. As any other observation of Cosmic Rays (CR), 
UHECR are characterized by three basic observables: spectrum, chemical composition and
anisotropy. In this short review we will mainly discuss the first two observables neglecting the 
discussion about a possible anisotropy in the arrival directions of UHECR.

The behavior of the UHECR spectrum is mainly conditioned by the interaction of such particles 
with the intervening astrophysical backgrounds and by the cosmological evolution of the Universe. 
In the energy range $E\simeq 10^{18} \div 10^{19}$ eV the propagation of UHE particles is extended 
over cosmological distances with a typical path length of the order of Gpc. Therefore one should also 
take into account the adiabatic energy losses suffered by particles because of the cosmological expansion 
of the Universe.

The background affecting the propagation of UHE protons is only the Cosmic Microwave Background (CMB). 
There are two spectral signatures that can be firmly related to the propagation of protons through this 
background: pair-production dip \cite{dip}, which is a rather faint feature caused by the pair production process:
\begin{equation}
p+\gamma_{CMB} \to e^{+} + e^{-} + p
\label{pair}
\end{equation}
and a sharp steepening of the spectrum caused by the pion photo-production:
\begin{equation}
p+\gamma_{CMB} \to \pi + p
\label{pion}
\end{equation}
called Greisen-Zatsepin-Kuzmin (GZK) cut-off \cite{GZK}. 

In the case of UHE nuclei the situation changes because, apart from CMB, also the Extragalactic Background Light 
(EBL) becomes relevant. The interaction processes that condition the propagation of UHE nuclei are pair production, 
that involves only the CMB background \cite{nuclei}, and photo-disintegration. The latter is the process in which a nucleus of 
atomic mass number $A$ because of the interaction with CMB and EBL looses one or more nucleons:
\begin{equation}
A+\gamma_{CMB,EBL} \to (A-nN) + nN
\label{disint}
\end{equation}
the most relevant process is the one nucleon emission $(n = 1)$ as discussed in \cite{nuclei}.The photo-disintegration 
of nuclei, together with the pair production process, produces a steepening in the observed spectrum. The exact position 
and behavior of the flux suppression depends on the nuclei species as well as on the details of the cosmological evolution 
of the EBL which, differently from CMB, is not known analytically being model dependent \cite{Stecker}. In general the nuclei 
flux steepening is not as sharp as the GZK and it is placed at lower energies respect to the GZK energy scale \cite{nuclei}.

The GZK cutoff is the most spectacular feature in the UHECR spectrum. However one must remember that it is defined 
only for UHE protons interacting with the CMB filed. The cutoff energy position is roughly fixed by the energy where the 
pair-production (Eq. (\ref{pair})) and the photo-pion production (Eq. (\ref{pion})) energy losses become equal, namely at
$E_{GZK} \simeq 50$ EeV \cite{Bere88}. The shape of the GZK flux suppression could also be model-depend: it depends 
on a possible local over-density or deficit of sources and can be mimicked by low enough values of the maximum energy 
that sources can provide. In contrast, the pair-production dip is practically model-independent mainly because protons
contributing to this energy range arrive from sources lying at cosmological distances.

From the experimental point of view the situation is far from being clear with different experiments claiming contradictory
results. The HiRes and, nowadays, the Telescope Array (TA) experiments show a proton dominated composition till the 
highest energies with a clear observation of the proton pair-production dip and GZK cut-off \cite{HiRes-TA}. Chemical 
composition observed by HiRes and TA is coherent with such picture showing a pure proton dominated spectrum 
starting from energies $E\simeq 10^{18}$ eV till the highest energies. The experimental picture changes taking into 
account the Auger observations. The spectrum observed by Auger shows a behavior not clearly understood in terms 
of the proton pair-production dip and GZK cut-off \cite{Auger}. This spectral behavior could be a signal of a substantial 
nuclei pollution in the spectrum, which is confirmed by the Auger observations on chemical composition that show 
a progressively heavy composition toward the highest energies, this tendency starts already at 
$E\gtrsim 4\times 10^{18}$ eV \cite{Auger}. 
In this short review we will discuss the main issues in the physics of UHECR presenting a twofold 
discussion from both experimental as well as theoretical point of view. The paper is organized as follows: 
in section \ref{dip-GZK} we will mainly discuss the two observable features connected with protons propagation, i.e. dip and 
GZK cut-off, presenting a comparison with HiRes and TA data, in section \ref{PAO} we will focus on the Auger observations 
of spectrum and chemical composition, discussing the consequences of a nuclei dominated spectrum, in section \ref{trans}
we will discuss the physics of the transition between galactic and extra-galactic CR, finally conclusions will take place in 
section \ref{conc}.

\section{HiRes and TA data: dip and GZK cut-off}
\label{dip-GZK}

\begin{figure}[ht]
\begin{center}
\includegraphics [width=0.3\textwidth]{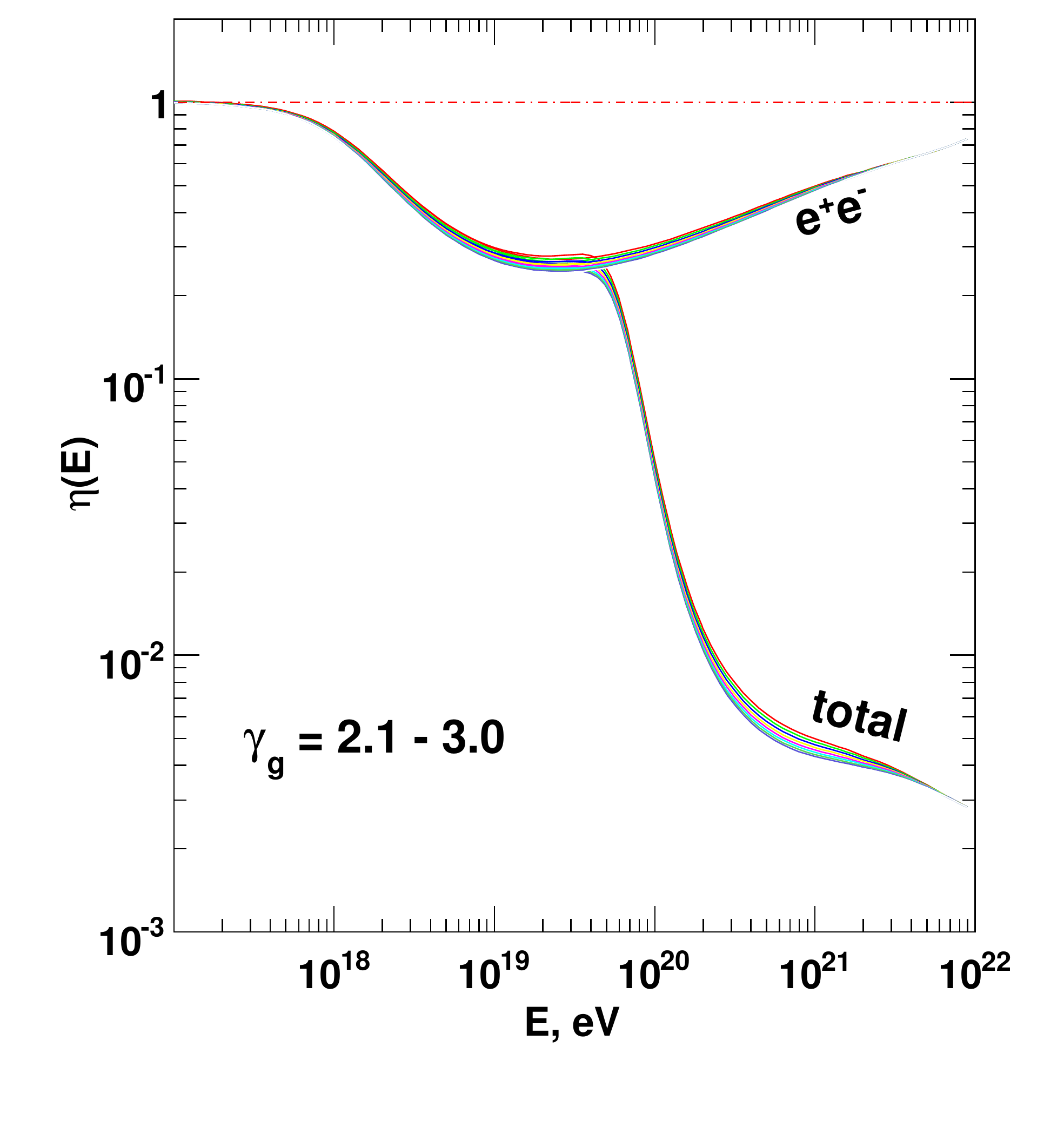}
\end{center}
\caption{Modification factor for different values of $\gamma_g$ (see text).}
\label{fig1}
\end{figure}

The dip is a quite faint feature that arises in the spectrum of UHE protons in the energy range $2\times 10^{18}$ eV
$\div5\times 10^{19}$ eV due to the pair production process; it is a unique imprint of the interaction of protons with 
the CMB background. The dip behavior is more pronounced when analyzed in terms of the modification factor 
$\eta(E)=J_p(E)/j_p^{umm}(E)$ \cite{dip}, defined as the ratio of proton spectrum divided by the so-called unmodified 
spectrum, where only adiabatic energy losses (expansion of the Universe) are taken into account. 

In this paper we will always assume a power law injection spectrum for UHECR, namely 
$Q(E_g)\propto E_{g}^{-\gamma_g}$
being $E_g$ the UHECR generation energy eventually bounded from above by the maximum attainable energy 
at the source $E^{max}_g$. A useful characteristic of the modification factor is its universality, it depends very weekly 
on the injection power law index and on various physical phenomena that can be included in the calculation \cite{dip}. 
In figure \ref{fig1} we plot $\eta(E)$ for different values of $\gamma_g$ (as labelled); it can be seen that if one includes in 
the calculation of $J_p$ only adiabatic energy losses then, by definition, $\eta(E)=1$ (dash dotted line). 
\begin{figure}[ht]
\begin{center}
\includegraphics [width=0.37\textwidth]{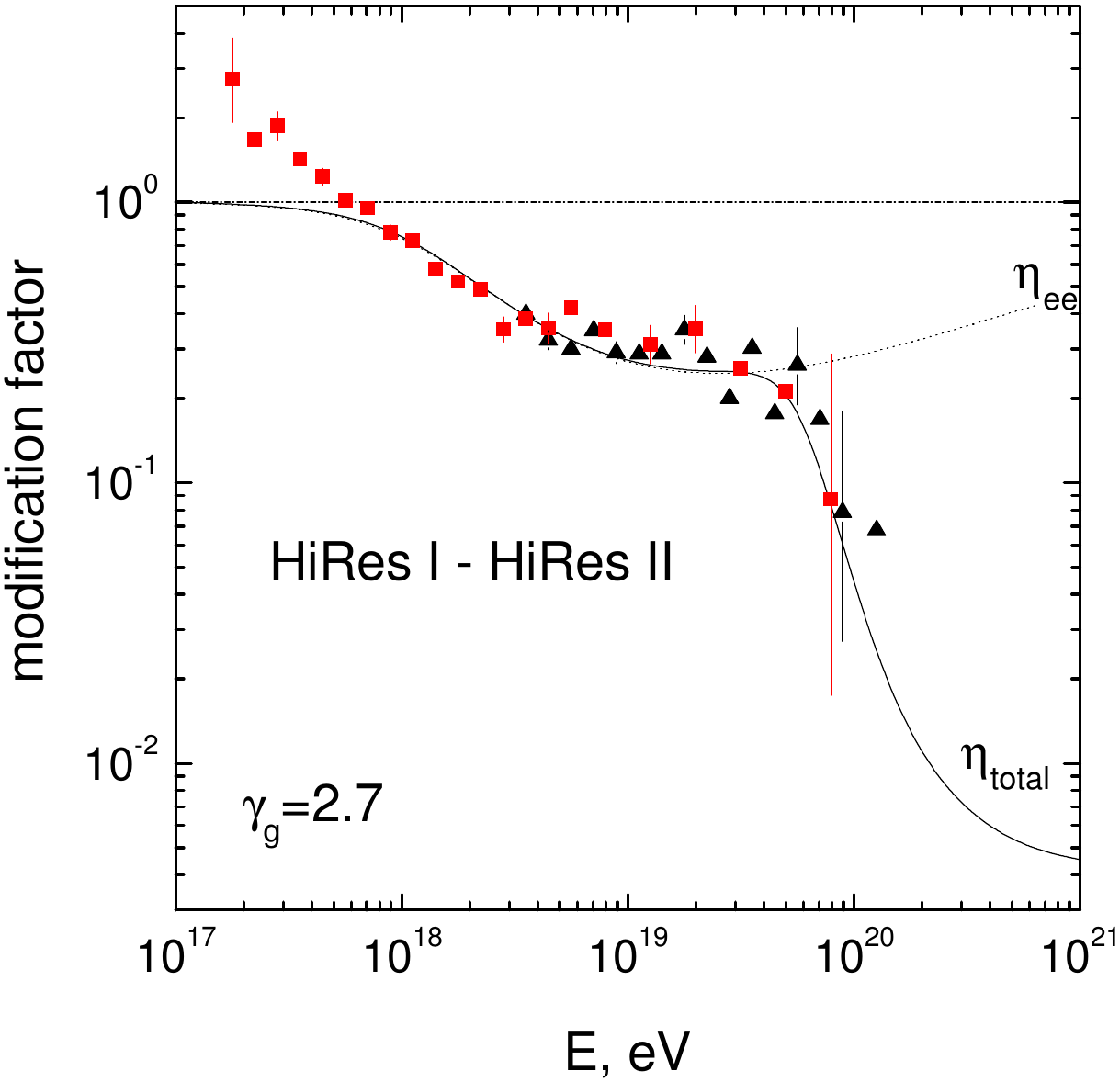}
\end{center}
\caption{The pair production dip compared with the HiRes experimental data 
\cite{HiRes-TA}, this experiment confirms the pair-production dip with good accuracy ($\chi^2/d.o.f. \simeq 1.0\div 1.2$).}
\label{fig2a}
\end{figure}
When $e^{+}e^{-}$ production is additionally included one obtains $\eta(E)$ as labelled in figure \ref{fig1}, this behavior
corresponds to the pair-production dip. Finally, if all channels of energy losses are taken into account $\eta(E)$ shows
also the GZK suppression, curve labelled 'total'. The only "critical" assumption at the very base of the dip behavior is the
proton dominated spectrum, already a nuclei pollution at the level of $15\%$ will spoil the behavior of figure \ref{fig1}. 

\begin{figure}[ht]
\begin{center}
\includegraphics [width=0.4\textwidth]{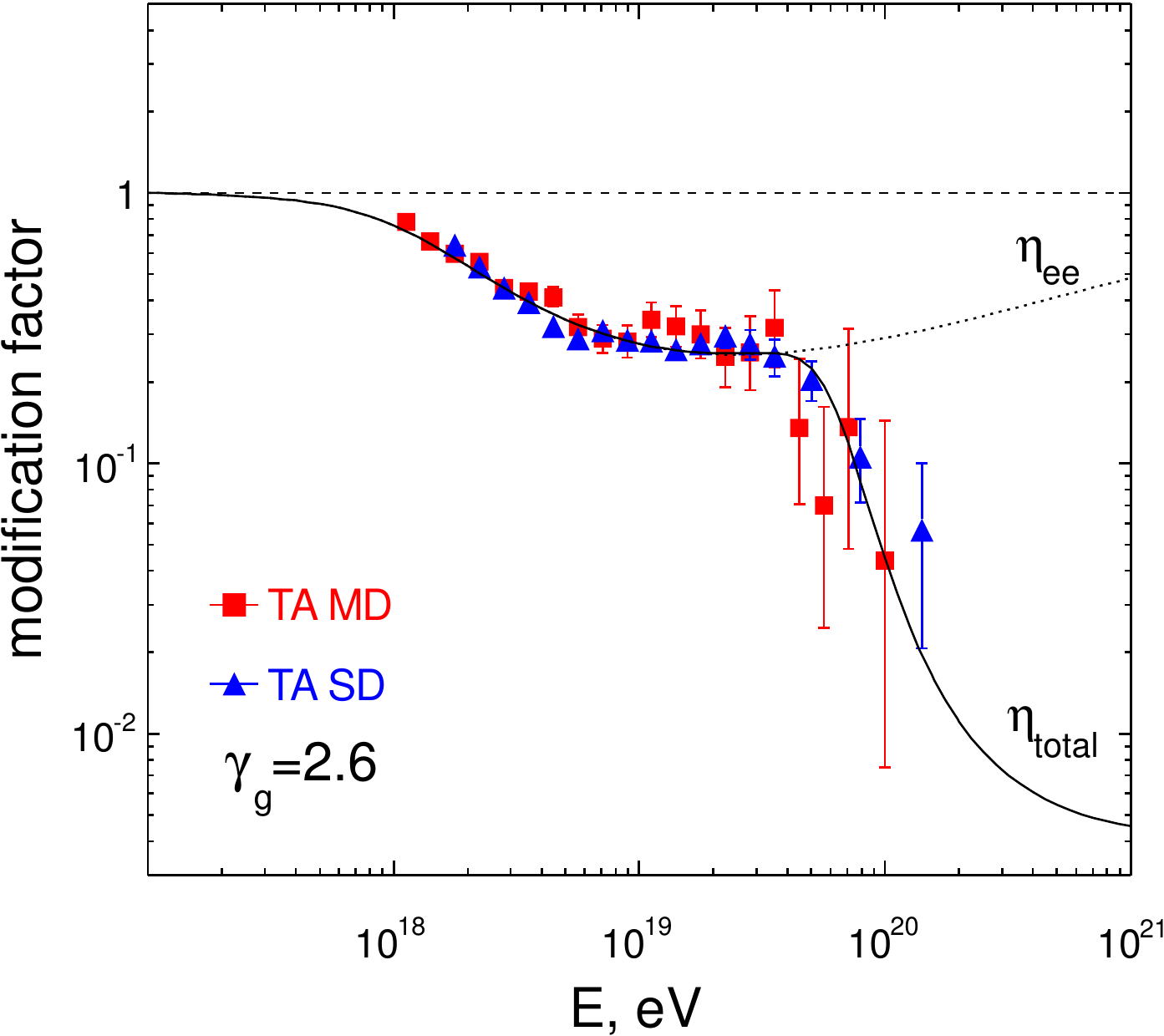}
\end{center}
\caption{The pair production dip compared with the TA experimental data \cite{HiRes-TA}.}
\label{fig2b}
\end{figure}

 Modification factor is a very useful tool to study the effect of the CMB radiation field on the propagation of UHE protons
and can be compared with observations simply dividing the observed spectra by the unmodified spectrum 
$J_{p}^{umm}(E)\propto E^{-\gamma_g}$. 
In figures \ref{fig2a},\ref{fig2b} we compare the modification factor with the 
observations of HiRes and TA \cite{HiRes-TA}. This comparison involves only two free parameters: the power 
law index at injection $\gamma_g$ and the total emissivity of UHE protons sources. 

The fit of HiRes and TA data to the pair-production dip is quite good with a $\chi^2/d.o.f\simeq 1$ and a best 
fit value of $\gamma_g\simeq 2.7$ for HiRes and $\gamma_g=2.6$ for TA.
From figures \ref{fig2a},\ref{fig2b} one can notice that the modification factor observed by HiRes exceeds the theoretical 
one around EeV energies; since by definition $\eta(E) \leqslant 1$ this excess signals the appearance of a new
component of CR at $E<1$ EeV. This component can be nothing else  but the galactic cosmic rays. Therefore the 
dip model for UHECR implies a transition from galactic to extra-galactic CR at $E\lesssim 1$ EeV, we will come back 
to this in section \ref{trans}.  

The experimental observations of HiRes and TA show a strong evidence of the GZK cut-off in the spectra, 
while Auger data, that we will discuss in the next session, seem less compatible with the expected GZK behavior
of the spectra. In general all experiments show a flux suppression at the highest energies, nevertheless to
ascribe it to the process of photo-pion production of protons on the CMB field (i.e. GZK suppression) one must 
prove that: (i) the energy scale of the cut-off and its shape correspond to the theoretical predictions, (ii) the observed 
chemical composition is strongly dominated by protons. 

The integral spectrum of protons $J_p(>E)$ has a power law behavior at low energy. Increasing the energy, 
because of the photo-pion production process, $J_p(>E)$ shows an abrupt suppression (i.e. GZK cut-off).
The energy scale at which this suppression arises can be evaluated as the energy at which the integral flux is reduced 
by half of its low energy value. This energy scale $E_{1/2}$ can be computed with extreme accuracy and it
is found to be practically model independent \cite{Bere88}, its theoretical prediction is 
$E_{1/2}=10^{19.72}~{\rm eV}\simeq 52.5$ EeV \cite{Bere88}. 

The HiRes and TA observations show a proton dominated spectrum at all energies $E>10^{18}$ eV \cite{HiRes-TA}, 
moreover the HiRes collaboration measured $E_{1/2}$ obtaining a value in a pretty good agreement with theoretical 
expectations, namely $E_{1/2}^{HiRes}=10^{19.73\pm 0.07}$ eV. These experimental evidences show a coherent 
picture of a proton dominated flux that exhibits the expected features: pair-production dip and photo-pion 
production suppression (GZK). 

\section{Auger data: heavy nuclei}
\label{PAO}

\begin{figure}[ht]
\begin{center}
\includegraphics [width=0.4\textwidth]{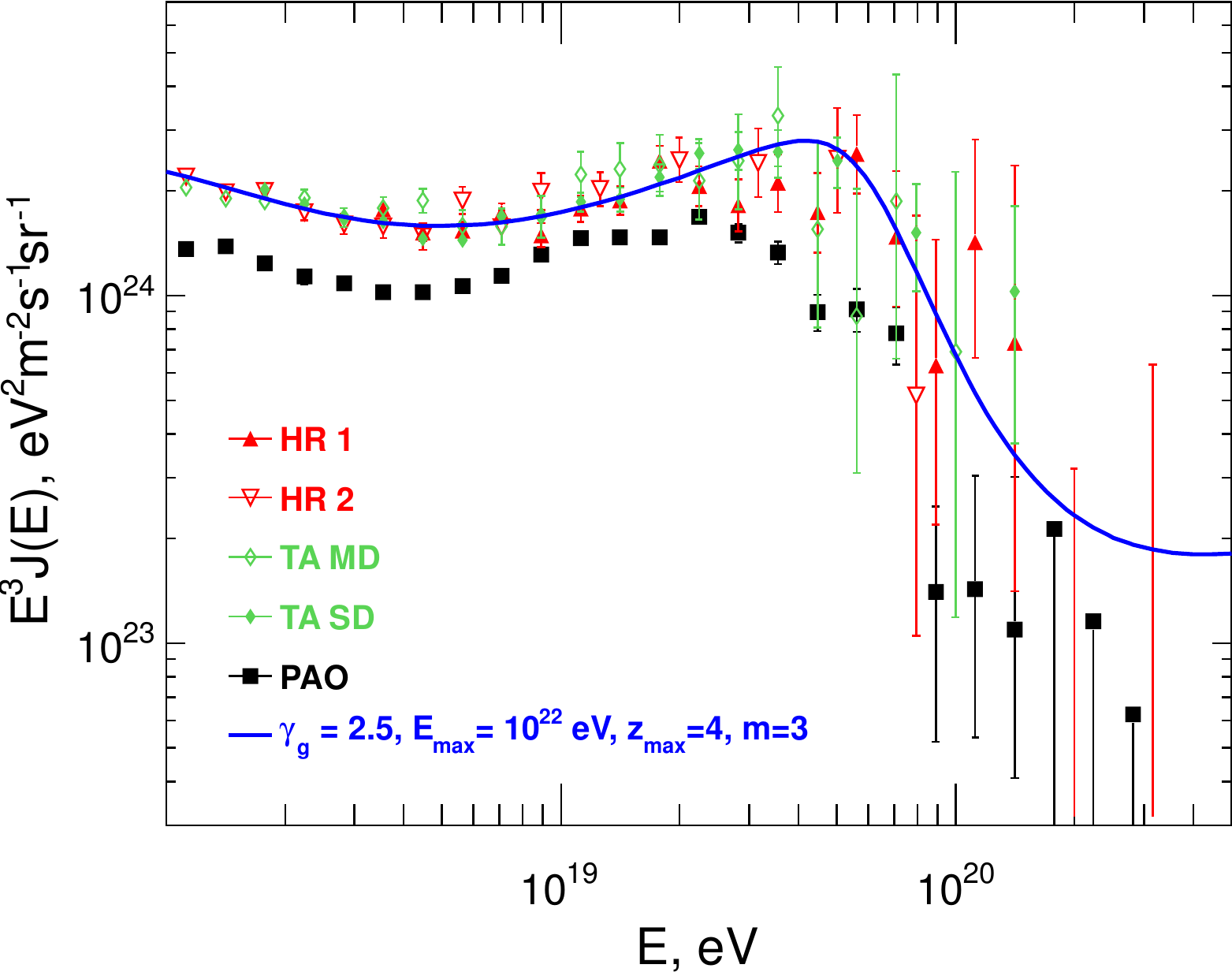}
\end{center}
\caption{Spectrum of UHECR as observed by HiRes, TA and Auger experiments (as labeled). 
Continuos line is the best fit of HiRes/TA with a pure proton composition (dip model)}
\label{fig3a}
\end{figure}

The spectrum observed by the Pierre Auger experiment, shown by filled squares in 
figure \ref{fig3a}, differs from the HiRes and TA spectra, shown by triangles and circles.  
On the other hand, as discussed in the previous section, the HiRes/TA data are well fitted by the pair-production 
dip, shown as continuos line in figure \ref{fig3a}.

In the framework of the dip model, taking into account the systematic errors in energy determination, 
one can try to shift the Auger energy scale having the dip behavior as reference. This procedure, 
proposed in \cite{dip}, is based on the fact that the energy position of the pair-production dip is rigidly 
fixed by the interaction of protons with the CMB radiation field, so that it can be used as a "standard candle". 
This approach is based on a single hypothesis: a pure proton composition of UHECR. 

\begin{figure}[ht]
\begin{center}
\includegraphics [width=0.4\textwidth]{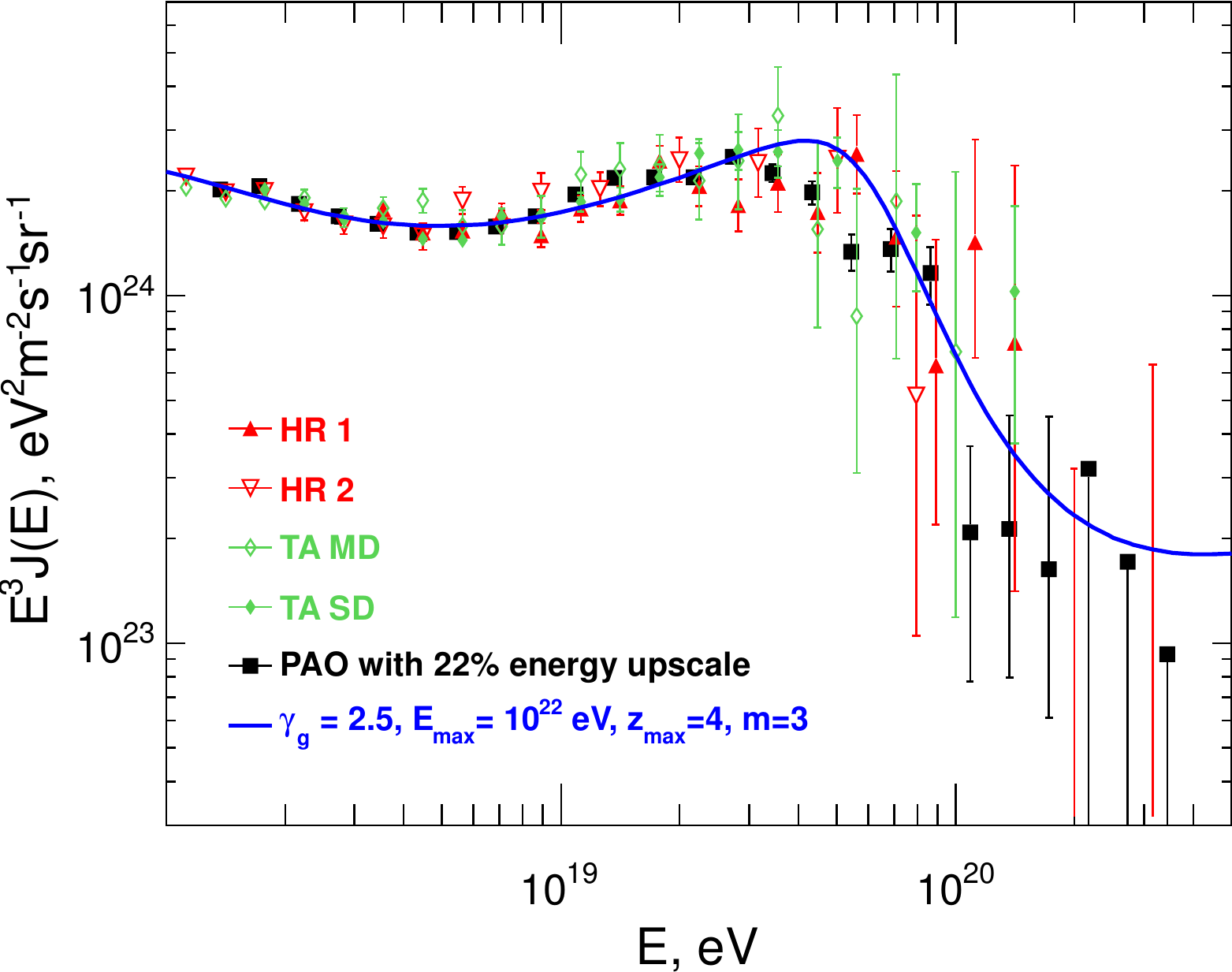}
\end{center}
\caption{The same as in \ref{fig3a} with the Auger spectrum shifted in energy as $E\to k_{cal}E$ with $k_{cal}=1.22$.}
\label{fig3b}
\end{figure}

Shifting accordingly to the dip "standard candle" the energy scale of the Auger data one has 
the result shown in figure \ref{fig3b}. Where the Auger data have been shifted according 
to $E\to k_{cal}E$ with $k_{cal}=1.22$, allowed by the systematic error claimed by the Auger collaboration 
\cite{Auger}. 

After the energy shift, the apparent coincidence of the Auger and HiRes/TA data is mainly related to the 
low energy part of the spectra (see figure \ref{fig3b}), while at the highest energies statistical 
uncertainties are too large to distinguish among spectra. Nevertheless, while HiRes and TA data are compatible with 
the GZK steepening the Auger data, which at energies around $50$ EeV still show low statistical 
errors, are not compatible with the behavior expected from the photo-pion production process of protons
on the CMB field. 

Even playing with the different parameters involved in the computation (i.e. injection power law index $\gamma_g$,
sources cosmological evolution, maximum acceleration energy, local over-density of sources, etc.) one 
could not reconcile the Auger spectrum behavior with the GZK expectation. Being the GZK steepening a clear 
signal of a proton dominated spectrum, the Auger spectrum is coherent with the chemical composition 
measured by this experiment at the highest energies which is quite incompatible with a pure proton composition. 
Auger data on chemical composition show a steady behavior that, starting 
already from energies around $3$ EeV, moves from a light (proton) to an heavier composition reaching an almost 
pure Iron composition at energies around $30$ EeV \cite{Auger}. These observations based on the Auger 
Fluorescence Detector (FD) are further straightened by the observations of the Auger Surface Detector (SD),
namely the shower muon content and the signal rise time in the Cherenkov tanks \cite{Auger-mu}.

The early steepening observed in the Auger spectrum can be easily explained in terms of nuclei propagation. 
Nevertheless, even assuming a rich chemical composition with different nuclei species injected at the source 
it is not easy to obtain a good explanation of both spectrum and chemical composition observed by Auger. 
A wide class of mixed composition models \cite{mixed}, while showing a good description of the spectrum, 
fail to fit the mass composition observed by Auger. Most of these models assume a mass composition 
similar to the one observed at galactic scales, therefore enhanced in protons, and the resulting mass 
composition starts to become heavier only at energies $E>50$ EeV \cite{mixed} where photo-pion 
production substantially depletes the proton component. 

At the galactic energy scales, the mass composition becoming heavier with increasing energy appears as 
a natural consequence of the rigidity dependent scenario for particles acceleration. The maximum 
acceleration energy that a single specie can reach is proportional to the particles charge 
$E^Z_{max}=Z E^p_{max}$ and the contribution to the spectrum at these energies of particles with charge
lower than $Z$ is suppressed. The disappointing model for UHECR \cite{disapp} was build exactly under such 
assumption on maximum energy. In \cite{disapp} was demonstrated that to avoid a proton dominated spectrum 
at the highest energies one must assume that the maximum energy for protons is in the range $4\div 10$ EeV. 
This conclusion remains valid for a large range of generation power law indexes $\gamma_g\simeq 2.0\div 2.8$, 
nevertheless to achieve a good description of the observed UHECR spectrum in the disappointing 
model one should assume a quite flat injection spectrum with $\gamma_g=2.0\div 2.2$.  
\begin{figure}[ht]
\begin{center}
\includegraphics [width=0.4\textwidth]{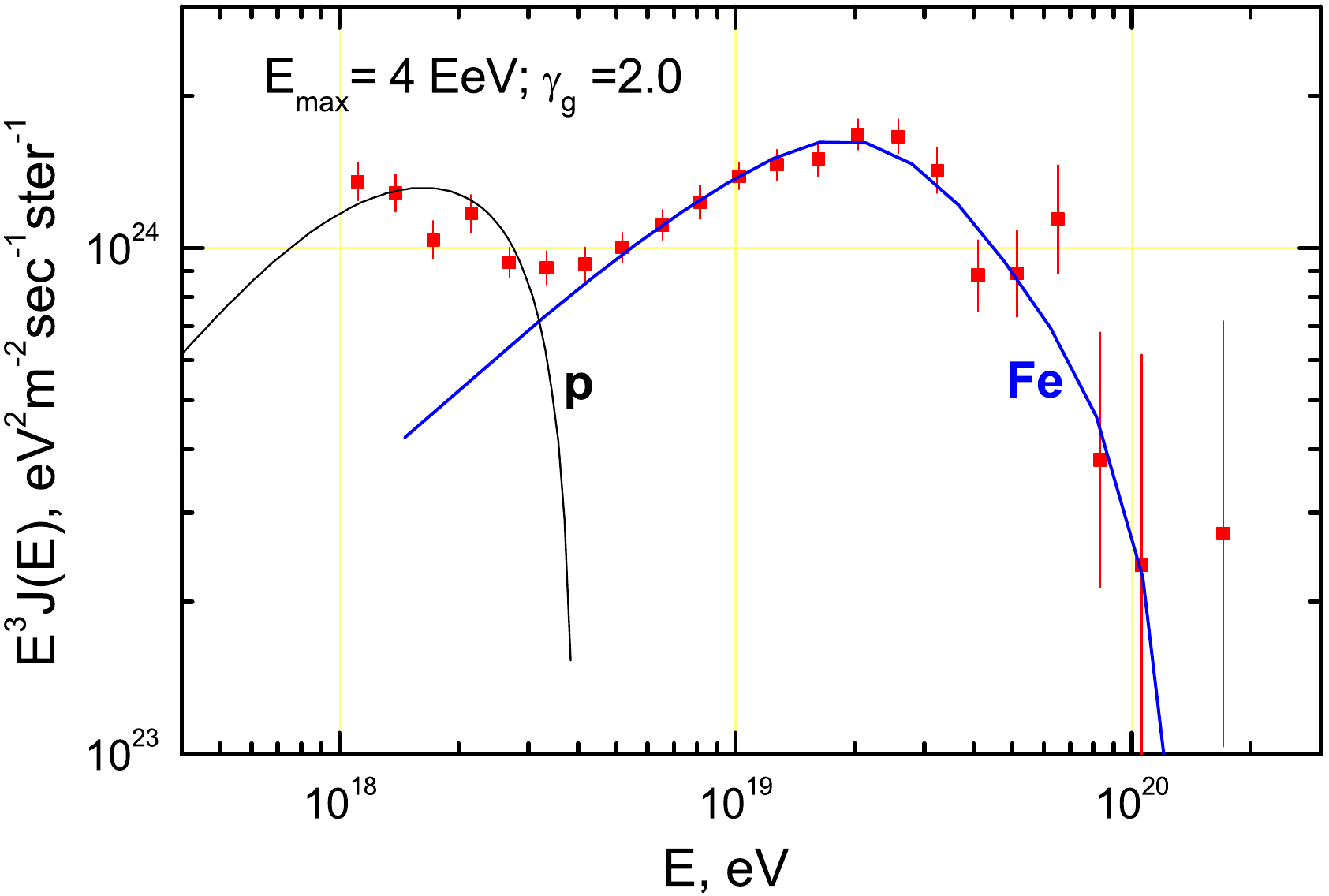}
\end{center}
\caption{Energy spectrum in the two components model with proton and iron nuclei for a homogeneous distribution
of sources with $\gamma_g=2.0$ and $E^p_{max}=4$ EeV.}
\label{fig4a}
\end{figure}
This fact, as we will discuss
in the next session, has important consequences on the transition between galactic and extra-galactic CR.
In figures \ref{fig4a},\ref{fig4b} we plot two different implementations of the disappointing model: a simple two component 
model with protons and iron (figure \ref{fig4a}), the same two components model with a diffusion cut-off in the iron
spectrum due to the possible presence of a $nG$ scale intergalactic magnetic field (figure \ref{fig4b}). 
Note that the gap in the spectrum in figure \ref{fig4b} can be filled by the presence of secondary nuclei produced by the
photo-disintegration of iron. 

This model was called "disappointing" because it corresponds to a lack of many signatures predicted in the alternative 
case of a proton dominated scenario, such as the cosmogenic neutrino production and correlation with astrophysical 
sources. The disappointing model as presented here and in \cite{disapp} is not complete because, while it reaches
a good description of the observed Auger spectrum, it does not try to describe also the chemical composition 
observed at all energies. 

The only successful theoretical attempt to fit at once spectrum and chemical composition of Auger was presented in 
\cite{Taylor} where it is assumed the presence of a very nearby source ($\simeq 70$ Mpc) with a very flat 
generation spectrum ($\gamma_g\simeq 1.6$) injecting heavy nuclei only.

\section{Galactic and extra-galactic cosmic rays}
\label{trans}

\begin{figure}[ht]
\begin{center}
\includegraphics [width=0.4\textwidth]{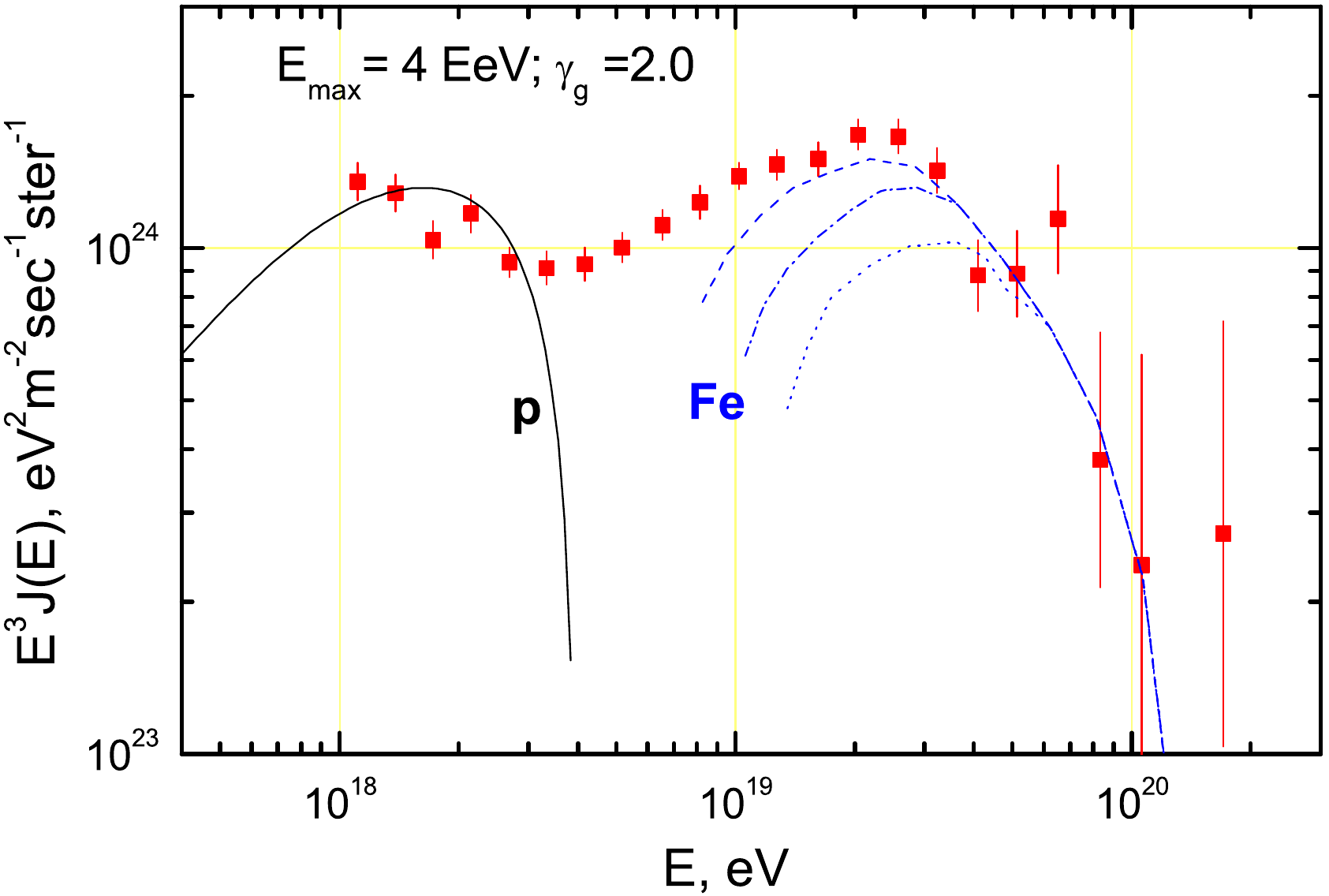}
\end{center}
\caption{As in figure \ref{fig4a} with a diffusion cut-off, the gap 
is expected to be filled by intermediate mass nuclei.}
\label{fig4b}
\end{figure}

The different models aiming to describe UHECR observations have different consequences when applied to
the transition from galactic to extra-galactic CR, therefore this theme of investigation has obtained an increasing 
interest in recent years. In general the physics of galactic CR is well understood in the framework of the standard 
model, which is based on the paradigm of Super Nova Remnant (SNR) as sources of galactic CR and Diffusive Shock 
Acceleration (DSA) as the mechanism of particles acceleration. In the standard model of galactic CR the observation of 
the knee at energy $2\times 10^{15}$ eV in the spectrum fixes the maximum energy of protons at the source 
and therefore, in a rigidity dependent scenario, the maximum energy for iron nuclei, the end of galactic CR, will be
around energies $10^{17}$ eV. 

\begin{figure}[ht]
\begin{center}
\includegraphics [width=0.4\textwidth]{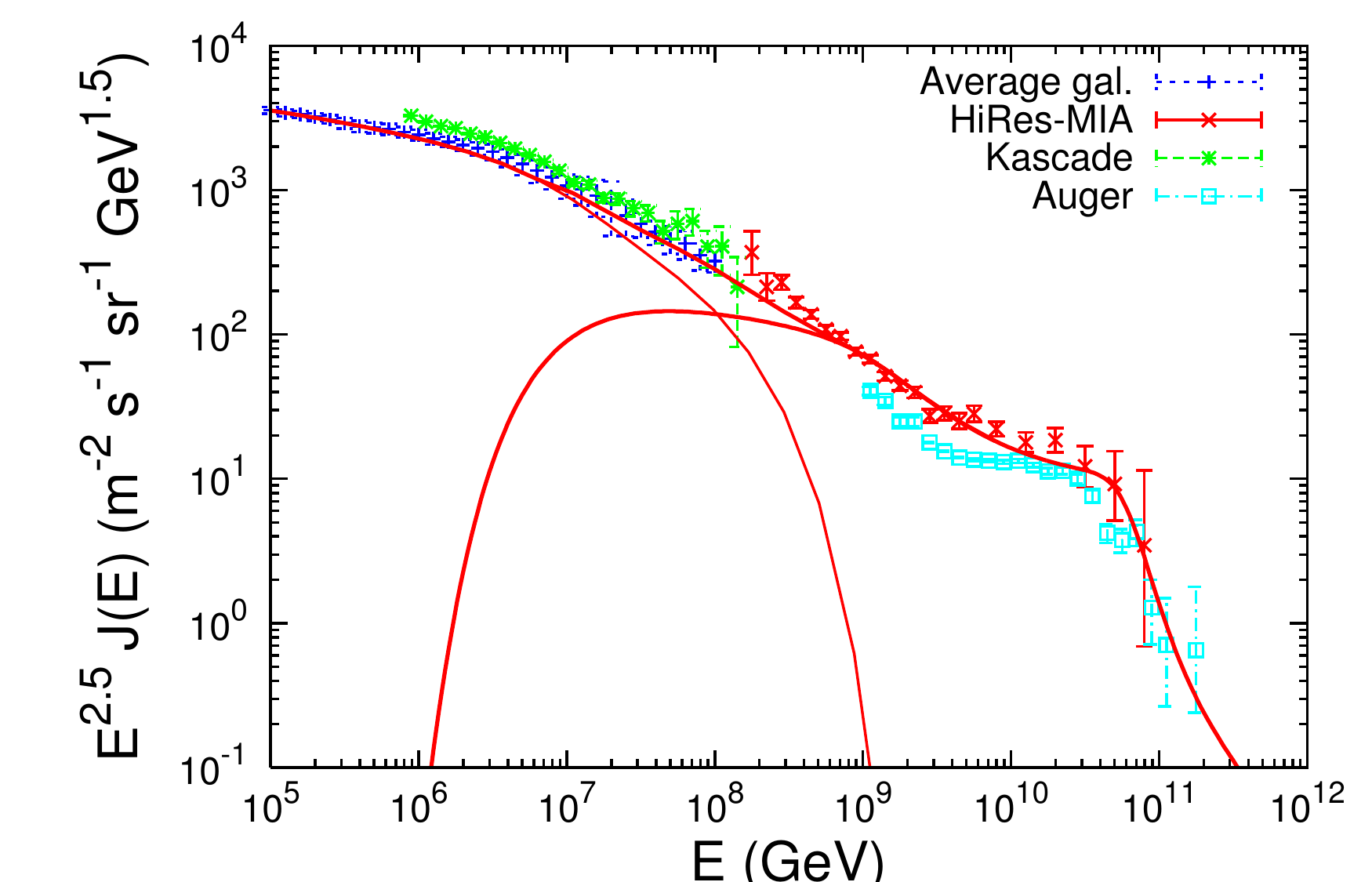}
\end{center}
\caption{Transition from galactic to extra-galactic CR in the case of the dip model, with 
$E_{max}=10^{21}$ eV and $\gamma_g=2.7$. Galactic CR flux is taken from 
\cite{PQ}, experimental data are from HiRes \cite{HiRes-TA}, Auger \cite{Auger}, Kascade \cite{Kascade} and 
the average over different experimental results as in \cite{average-gal}.}
\label{fig5a}
\end{figure}

\begin{figure}[ht]
\begin{center}
\includegraphics [width=0.4\textwidth]{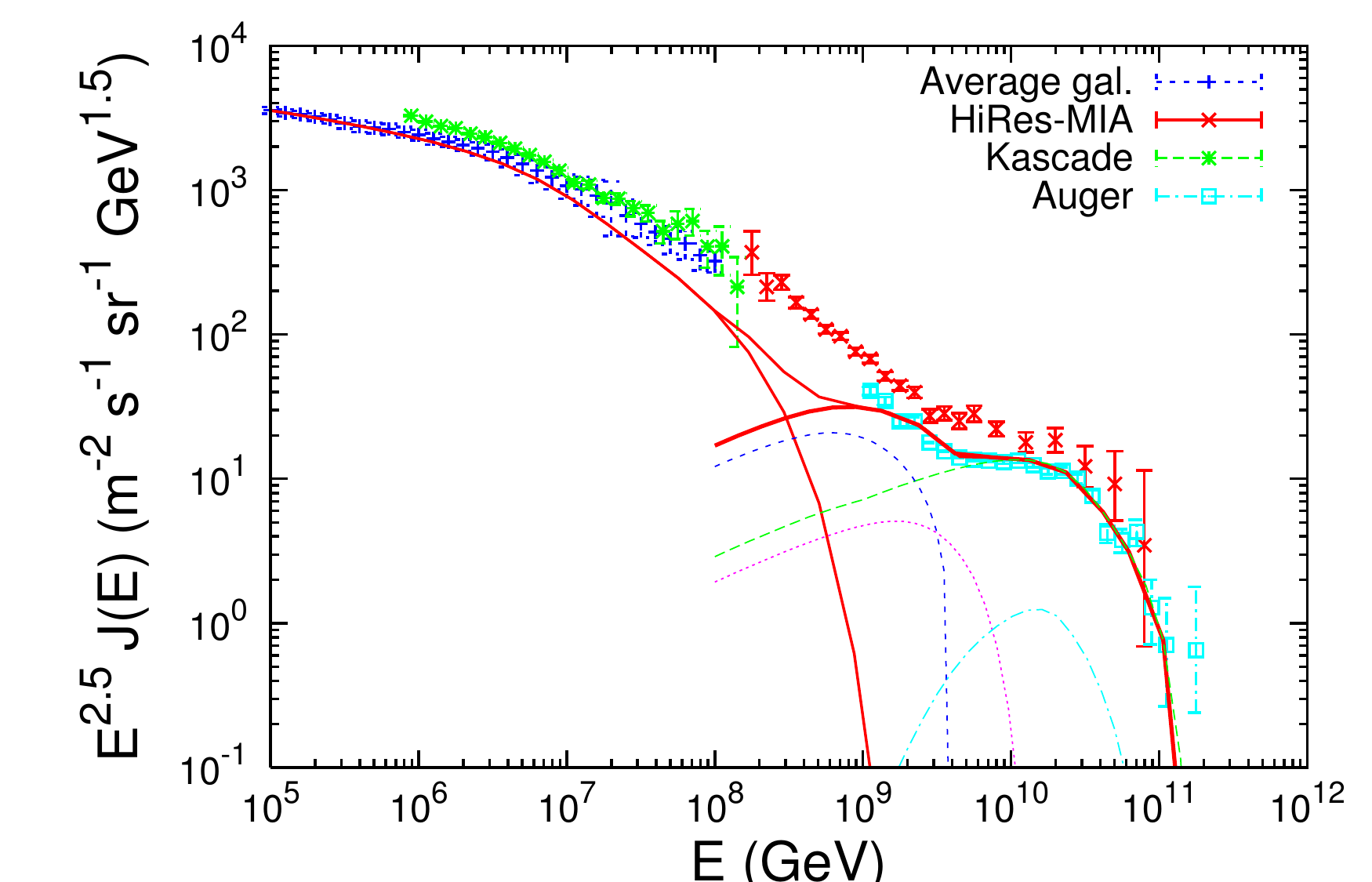}
\end{center}
\caption{Transition from galactic to extra-galactic CR in the 
case of the disappointing model assuming a mixture of protons, He and Fe nuclei at the source with 
$E_{max}^p=4\times 10^{18}$ eV and $\gamma_g=2.0$. Galactic CR flux as in figure \ref{fig5a}.}
\label{fig5b}
\end{figure}

In figures \ref{fig5a},\ref{fig5b} we plot together the fluxes of galactic and extra-galactic CR assuming the two different models 
dip (figure \ref{fig5a}) and disappointing (figure \ref{fig5b}) for UHECR and using the flux of galactic CR as computed 
in \cite{PQ}, that takes into account the scale distribution of SNR in the galaxy. The experimental data of figures 
\ref{fig5a},\ref{fig5b} are those of HiRes \cite{HiRes-TA} and Auger \cite{Auger} on the UHECR side and, on the galactic 
side, the data of all particles spectrum of Kascade \cite{Kascade} and an average of the fluxes measured by different
experiments as presented in \cite{average-gal}. 

The matching of galactic and extragalactic fluxes in the case of the dip model (figure \ref{fig5a}) gives a very 
good description of the experimental data, reproducing in an extremely accurate way the spectra observations in the 
intermediate energetic regime where the transition is supposed to stand. The case of the disappointing model gives 
a less accurate description of the experimental data (figure \ref{fig5b}) with a slightly suppression of the
theoretical flux in the transition region not seen experimentally. 
The less good agreement with experimental data obtained, in the transition region, with the disappointing model respect 
to the dip model is a direct consequence of two important facts: (i) the quite flat power law injection index 
($\gamma_g=2.0$) of the disappointing model, (ii) the quite low maximum energy associated to iron nuclei 
($E_{max}^{Fe}\simeq 10^{17}$ eV) in the standard model of galactic CR. These two facts produce a lacking of 
particles in the transition region not seen experimentally, to avoid such circumstance one should assume a larger
maximum energy for galactic iron. Only in this way in-fact it is possible to fill the gap in the transition region making 
the theoretical spectrum compatible with observations. 

Let us conclude this section emphasizing the importance of the study of the transition between galactic and 
extra-galactic CR. In general, the transition is the superposition of the low energy tail of UHECR with the high energy 
tail of galactic CR and it is supposed to stand in the energy range $0.1\div 10$ EeV. As discussed above the
informations obtained at these energies on galactic CR involve the maximum acceleration energy of particles and their
chemical composition. Therefore, assuming the standard model of galactic CR, the study of the transition can unveil the
whole picture of the origin of galactic CR. Moreover, the low energy tail of UHECR can give key informations on the
(possible) existence of the pair-production dip and on the propagation of UHECR in intergalactic magnetic fields
\cite{review}. Therefore, the experimental studies in the transition region are of paramount importance in this field 
of research, with the mass composition measurements being probably the most important task \cite{review}.  

 \section{Discussion and conclusions}
\label{conc}
We face nowadays the most serious disagreement in the observations of UHECR between the data of Auger and 
HiRes/TA. While Auger points toward an heavy composition at the highest energies, HiRes/TA  shows a proton 
dominated composition at all energies. 

The most self-consistent conclusions on the nature of UHECR are obtained at present by HiRes/TA: they observe 
a proton dominated chemical composition with a spectrum that shows both expected signatures of protons, i.e. the 
pair-production dip and the GZK cut-off. Moreover, in the HiRes data the GZK cut-off is also found in the integral
spectrum with the expected value of the $E_{1/2}$ energy scale. The picture emerging from HiRes/TA data confirms 
the dip model \cite{dip} and places the transition between galactic and extra-galactic CR at energies around 
$10^{18}$ eV in very good agreement with galactic CR observations. The only discrepant feature of the HiRes/TA 
observations consists in the absolute lacking of any anisotropy signal, expected at the highest energies in the case of 
a pure proton composition. 

The Auger data are quite different. Measurement of chemical composition shows a steadily increasing mass starting
from energies around $3$ EeV, observation recently confirmed by independent data on muon content in the showers
\cite{Auger-mu}. The Auger chemical composition excludes the dip model and the observed high energy steepening 
as the GZK cut-off. At present there is no nuclei based model which explains simultaneously the Auger energy spectrum
and chemical composition, apart  from the model proposed in \cite{Taylor} based on an unlikely framework with a nearby
source ($70$ Mpc) that injects only heavy nuclei with a very flat spectrum ($\gamma_g=1.6$). 

\begin{figure}[ht]
\begin{center}
\includegraphics [width=0.4\textwidth]{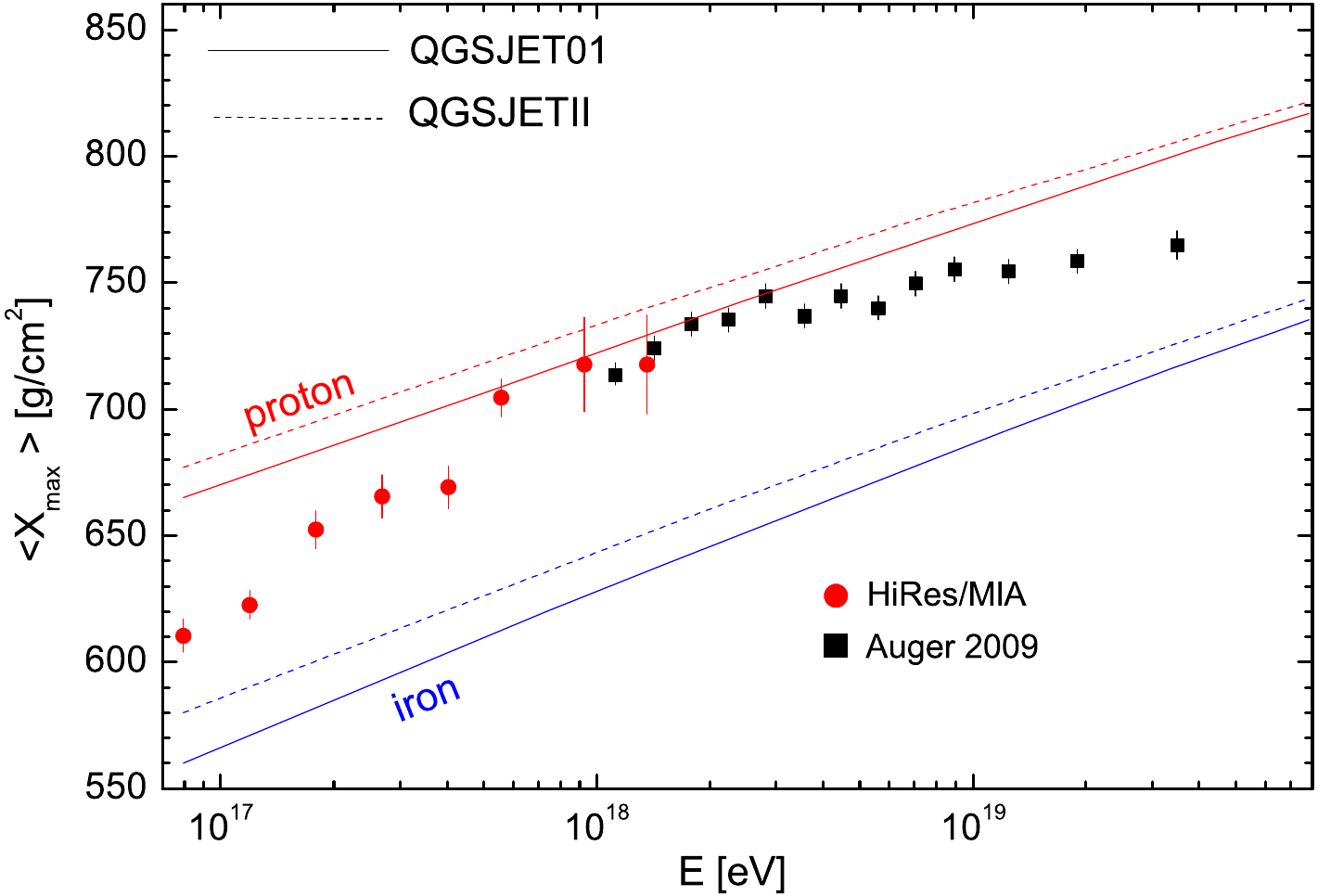}
\end{center}
\caption{Elongation curve as measured by Auger \cite{Auger}. The calculated reference values for proton and iron 
are those computed in the framework of the QGSJET1-2 model \cite{QGSJET}.}
\label{fig6a}
\end{figure}

We can conclude that the key issue to distinguish among different models in UHECR physics is related to the 
measurement of chemical composition. The best method at present to determine the UHECR composition is given 
by the measure of the position of the maximum of the cascade developed in the atmosphere, i.e. the elongation curve 
$\langle X_{max} \rangle(E)$. In figure \ref{fig6a} we plot the elongation curve as observed by Auger
\cite{Auger} and in figure \ref{fig6b} the same quantity measured by HiRes \cite{HiRes-TA}. 
The calculated reference values for proton and iron plotted in figures \ref{fig6a},\ref{fig6b} are those computed in the
framework of the QGSJET01 and QGSJET02 models \cite{QGSJET}. 

Unfortunately, the determination of chemical composition through $\langle X_{max} \rangle (E)$ suffers from many
systematics due to the experimental approach and uncertainties in the interaction model. 
Systematic errors in the $\langle X_{max} \rangle$ measurements can be as large as $20\div 25$ g/cm$^2$, to be
compared with the difference of about $100$ g/cm$^2$ between $\langle X_{max} \rangle$ of proton and iron. A better
sensitivity to distinguish different nuclei is given by the width of the $X_{max}$ distribution, i.e. $RMS(X_{max})$
\cite{RMS}.

We conclude stating that renewed experimental efforts are needed to reach a more reliable determination of the chemical 
composition of UHECR solving the apparent contradiction in the Auger and HiRes/TA data. 

\section*{Aknowledgements}
I'm grateful to V. Berezinsky, P. Blasi, A. Gazizov and S. Grigorieva for our joint activity in the field of UHECR physics.

\end{document}